\documentclass[11pt,letterpaper]{article}

\usepackage{graphicx}

\begin{document}

\def\beq{\begin{equation}}
\def\eeq{\end{equation}}
\def\bea{\begin{eqnarray}}
\def\eea{\end{eqnarray}}

\def\a{\alpha}
\def\r{\rho}
\def\s{\sigma}
\def\m{\mu}
\def\n{\nu}
\def\k{\kappa}
\def\g{\gamma}
\def\L{\Lambda}
\def\D{\Delta}
\def\la{\langle}
\def\ra{\rangle}
\def\o{\omega}
\def\d{\delta}
\def\p{\partial}
\def\Se{$S_E$ }
\def\Sa{$S_{\rm atmo}$ }

\def\tphi{\tilde{\phi}}
\def\tu{\tilde{u}}
\def\hv{\hat{v}}

\def\nab{\nabla}
\def\del{\partial}

\def\half{\textstyle{\frac{1}{2}}}
\def\quarter{\textstyle{\frac{1}{4}}}

\begin{center} {\Large \bf
Horizon surface gravity as 2d geodesic expansion}
\end{center}

\vskip 5mm
\begin{center} \large
{Ted Jacobson$^{*}$\footnote{E-mail: jacobson@umd.edu} and Renaud Parentani$^{\dagger}$\footnote{E-mail: Renaud.Parentani@th.u-psud.fr}}\end{center}

\vskip  0.5 cm
{\centerline{$^{*}$Department of Physics}}
{\centerline{\it University of Maryland}} 
{\centerline{College Park, MD 20742-4111, USA}}

\vskip  0.5 cm
{\centerline{$^{\dagger}$Laboratoire de 
Physique Th\'eorique}}
{\centerline{\it CNRS UMR 8627,
Universit\'e Paris-Sud 11}} 
{\centerline{91405 Orsay Cedex,
France}}

\begin{abstract}
The surface gravity of any Killing horizon,
in any spacetime dimension, can be interpreted
as a local, two-dimensional expansion rate seen by freely falling
observers when they cross the horizon.
Any two-dimensional 
congruence of geodesics invariant
under the Killing flow can be used to define this expansion,
provided that the
observers have unit Killing energy. 

\end{abstract}


\section{Introduction}

A Killing horizon is a null hypersurface 
whose null generators are flow lines
of a Killing field.
At each point of a Killing horizon, a
scalar quantity 
called the surface gravity is
defined. Its usual definition refers only to the
properties of the Killing vector (and the metric)~\cite{geometry}. 
The main purpose of this note is to indicate how it can
be defined instead as the rate of expansion of space
in the direction of the Killing frame
as viewed by 
freely falling observers
crossing the
horizon. This is analogous to cosmological expansion,
but it involves only one spatial 
direction, 
and no preferred observer is selected. Any %
freely falling
observer
with unit Killing energy will determine the same
expansion
at the horizon,
which is equal to the surface gravity.

This investigation was initially motivated by the
expression for surface gravity in Painlev\'{e}-Gullstrand
type coordinates. Consider for example the two dimensional
line element
\beq\label{PG}
ds^2 =  dt^2 - (dx - v(x)dt)^2,
\eeq
which has a Killing vector $\partial_t$ and a
Killing horizon where $v(x)=\pm 1$.
The surface gravity turns out to be given by the
gradient $dv/dx$
evaluated at the horizon. 
This may be interpreted as the fractional rate of 
expansion of the flow defined by the geodesics 
satisfying $dx=v dt$, in the following sense. 
These geodesics are orthogonal to the 
constant $t$ lines, on which $dx$ measures proper 
length. The proper distance $\d x$ between neighboring 
geodesics therefore satisfies $d(\ln \d x)/dt = \d v/\d x = dv/dx$.
Since the proper time along these
geodesics is just $dt$, this is the fractional rate of 
change with respect to proper time. 
Here we formulate what amounts to a covariant version
of this relation between surface gravity and two-dimensional
expansion,
and we demonstrate
its universal applicability to any Killing horizon in any spacetime
dimension, with the observer moving in any direction.

Aside from the 
geometric
perspective it offers,
another motivation for this note
is that the 2d expansion field,
defined both on and off the horizon,
may turn out to be physically 
relevant in settings where a local preferred frame is 
present. Condensed matter
black hole analogs~\cite{Unruh:1980cg, Jacobson:1999zk,
Barcelo:2005fc},  theories with
a dynamical 
``aether"~\cite{Jacobson:2008aj,Mukohyama:2005rw,Dubovsky:2007zi}, 
and Lorentz violating dispersion~\cite{Jacobson:1999zk} 
or dissipation~\cite{Parentani:2007uq} 
are examples of such settings.

\section{Surface gravity}

There are several ways to define
the
surface gravity $\kappa$ using
only the Killing vector $\chi^a$ and the metric.
For the present
purposes, the most convenient is via
the relation
\beq\label{kappa}
 \nab_a\chi^2 = -2 \k \chi_a, 
 \eeq
evaluated at the horizon.
The gradient of $\chi^2$
 is normal to the horizon, since
 this function is constant (equal to zero) on the horizon.
 The Killing vector is
 also normal to the horizon, since it lies along the null
 tangent direction which is always the normal direction
 to a null hypersurface.
 These vectors are therefore parallel,
so $\k$ is well-defined by
 (\ref{kappa}). Note however that the value of $\k$
 depends on the normalization of the Killing vector,
 which is not determined by the symmetry alone.
 Typically the norm is fixed to be unity
 on the worldline of some Killing observer, for example
 at infinity for an asymptotically flat black hole spacetime.
 
 We can use (\ref{kappa}) to find the
 surface gravity for the metric (\ref{PG}) 
 with respect to the  Killing vector $\del_t$.
 (In this example the
Killing vector has unit norm where $v(x)=0$.)
The $x$ component of (\ref{kappa}) yields
 $\k = (-1/2)g_{tt,x}/g_{xt} = v_{,x}$.

 \section{2d expansion}

The expansion of a congruence of curves is defined
by the divergence $\theta=\nabla_a u^a$ of the unit
tangent vector field $u^a$, where $\nab_a$ is the covariant
derivative operator. Here we will apply this notion
to two-dimensional timelike congruences 
in a spacetime of any dimension.
The 2d surface generated by such a congruence
has an induced
metric and induced covariant
derivative operator $D_a$,
obtained from $\nabla_a$ by restricting the
derivative to directions in the surface and projecting
the Levi-Civita connection onto the surface.
In terms of the surface
projector $h^a{}_b$,  the 2d expansion is given by
\beq\label{theta2d}
\theta_{2d} \equiv  D_a u^a = h^a{}_b \, \nabla_a u^b.
\eeq

The 2d congruences of interest here are composed
of timelike geodesics
and are invariant under the flow of a Killing vector
$\chi^a$. A timelike vector $u^a$ at
a single point $p$ uniquely determines such a congruence,
namely, the geodesic through $p$ with tangent $u^a$,
together with the image of this geodesic along the Killing
flow. The Lie derivative ${\cal L}_\chi u = [\chi,u]$
of the resulting 2d $u^a$ field
vanishes by construction. Thus
the derivatives of $u$ and $\chi$ are related by 
\beq\label{Lie}
\chi^a \nab_a u^b
=u^a\nab_a \chi^b.
\eeq
Note that the congruence defined in this way,
given $u^a$ at a single point, is independent of the
normalization of the Killing vector.

The tangent plane to the 2d congruence
at each point is spanned by $u^a$ and an orthogonal,
unit spacelike vector $s^a$ 
in terms of which
the orthogonal projector is 
simply
given by
\beq\label{projectorus}
h^a{}_b = u^au_b - s^a s_b.\eeq
Expressing $s^a$ in terms of $u^a$ and $\chi^a$ as
\beq\label{s}
s^a = (\chi\cdot s)^{-1}((\chi\cdot u)\, u^a - \chi^a),
\eeq
and 
inserting (\ref{projectorus})
into the expression (\ref{theta2d}), 
the terms involving contractions with $u$
vanish either because of the geodesic equation
or because of the normalization
of $u$, leaving just
\beq\label{theta2dchi}
\theta_{2d} = -(\chi\cdot s)^{-2} \chi_b\,  \chi^a\nabla_a\,  u^b.
\eeq
Using the Lie dragging condition (\ref{Lie}),
and $\chi^2=(\chi\cdot u)^2 -(\chi\cdot s)^2$,
(\ref{theta2dchi}) becomes
\beq\label{Du}
\theta_{2d}  =\frac{\half u^a\nab_a\chi^2}{\chi^2-(\chi\cdot u)^2}.
\eeq
This expression for the expansion
holds for the 2d
geodesic
Killing
congruence generated by any unit timelike vector.
It is independent of the overall scale of the
Killing vector, and its
value generally depends on the direction
of the unit vector.

It is worth noting that 
the 2d expansion is also equal to
the spatial derivative of the velocity
relative to the Killing frame, 
generalizing the 
coordinate expression $dv/dx$ 
for the expansion of the flow in (\ref{PG}). 
Indeed, using the decomposition
\beq\label{v}
\chi^a = (\chi\cdot u)(u^a - v s^a),
\eeq
where 
$ v = (\chi\cdot s)/(\chi\cdot u) $ 
is
the velocity of $u^a$ relative to
the Killing frame, (\ref{theta2d}) becomes
\bea\label{s.v}
\theta_{2d} 
&=& - s^a s_b \nabla_a u^b   
\nonumber\\ 
&=& - s^a s_b \nabla_a((u\cdot\chi)^{-1}\chi^b + v s^b)
\nonumber\\
&=&  s^a\nabla_a\, v.
\eea
(When $u^a$ is chosen tangent to the free fall trajectory $dx=v(x)dt$ 
in the coordinate system of (\ref{PG}), $v$ in (\ref{v}) coincides with 
$v(x)$, and the final expression in (\ref{s.v}) reads $dv/dx$.)
In the last 
step we used 
(i) the fact that the derivative of $u\cdot\chi$ along $s^a$ vanishes
(since it is constant along each geodesic 
and along the Killing flow), (ii) Killing's
equation $\nabla_{(a} \chi_{b)}=0$, and (iii) $s_b s^b=-1$.
Moreover, using (\ref{v}) and the Killing symmetry $\chi^a\nabla_a v=0$,
this result may be expressed in terms of the fractional rate of
change of $v$
respect to proper time, 
\beq
\theta_{2d}  = u^a\nabla_a \ln v.
\eeq

\section{2d expansion and surface gravity}

On the horizon, $\chi^2$ in the denominator of (\ref{Du}) vanishes,
hence
\beq\label{2dthetahor1}
\theta_{2d}^{\rm hor}  = -\frac{\half u^a\nab_a\chi^2}{(\chi\cdot u)^2}.
\eeq
Using (\ref{kappa}) this yields
\beq \label{kappatheta}
\kappa=(u\cdot\chi)\, \theta_{2d}^{\rm hor}.
\eeq
The 2d expansion on the horizon is thus
equal to the surface gravity, provided that
$\chi^a$ is normalized such that
$u^a$ has unit Killing energy,
i.e.\  $u\cdot\chi=1$.
It is curious that all dependence on $u^a$ except for
the Killing energy drops out of the 2d expansion when
evaluated on the horizon.

If a normalization of the
Killing vector is fixed, then at each point of the
horizon there is a  
($D-2$)-parameter family
of 
unit timelike 
$u^a$ with unit Killing energy,
$D$ being the dimension of space-time.
For all of the corresponding 
observers the 2d expansion
is equal to the surface gravity. Alternatively,
for {\it any} observer, if the Killing field is normalized
so the observer has unit Killing energy,
the surface gravity is equal to the 2d expansion.

\section{Examples}

We now illustrate the central result
of this note with four
examples. First, for an asymptotically flat rotating
black hole in four dimensions,
the horizon-generating Killing vector is
$\chi= \partial_t+\Omega_H\partial_\phi$, where
$\partial_t$ is the time-translation
and $\partial_\phi$ is the axial rotation.
If $\partial_t$ is normalized to unity at
spatial infinity as usual, then
geodesics that fall from rest at infinity with zero 
angular momentum have
$u\cdot\partial_t=1$ and  $u\cdot\partial_\phi=0$, so
they have unit Killing energy, $u\cdot\chi=1$.
When they fall across the horizon such observers
measure a 2d expansion that is precisely
the surface gravity of $\chi$. But these are not
the only observers with unit Killing energy.
At each point
there is a two-parameter family of such observers,
all of whom
measure the same 2d expansion at the horizon for their
associated geodesic Killing congruence.
The 4-velocities of these observers are
related to each other by
Lorentz transformations (in the tangent space)
that leave the Killing vector fixed. On the
horizon these are null rotations, while outside they
are spatial rotations or boosts, depending on
whether $\chi$ is timelike or spacelike.

\begin{figure}
\begin{center}
 \includegraphics[angle=0,width=4.4cm]{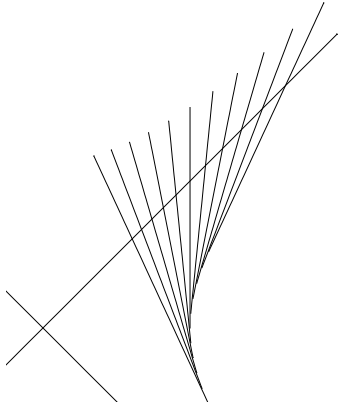}
 \end{center}
\caption{\label{ff} \small Geodesic congruence in 
Minkowski spacetime invariant under a boost
Killing flow.
The Killing (Rindler) horizon is the
pair of perpendicular lines. The geodesic segments 
are straight lines launched tangent to a hyperbola.
The expansion of the congruence at horizon crossing is
the reciprocal of the proper distance
from the hyperbola to the bifurcation point (where the
horizon lines cross);
it is also equal to the surface gravity 
divided by the 
Killing energy of the geodesics.}
\end{figure}

Our second example is the Rindler horizon in 2d
Minkowski spacetime, generated by the boost Killing
vector $x\partial_t+t\partial_x$, where $t$ and $x$
are Minkowski coordinates.  Let us consider the
geodesic Killing congruence determined by the unit
timelike vector $u=\partial_t$ located at the point
$(t,x)=(0,x_0)$. The fiducial geodesic through this
point with this tangent vector is a straight line in
the $t$ direction, while the geodesics obtained by
dragging along the Killing flow are straight lines
tangent at the other points along the hyperbola $x^2-t^2 =
x_0^2$ (see Fig.~\ref{ff}).
For this congruence we have
$\theta_{2d}^{\rm hor}=  
\del_t u^t + \del_x u^x$,
which at the horizon-crossing point 
$(x_0,x_0)$ evaluates to
$\theta_{2d}^{\rm hor}= 1/x_0$.
The Killing energy of this congruence is
given by
$u\cdot(x\partial_t+t\partial_x)=x_0$,
so according to
(\ref{kappatheta}) the surface gravity
is then $\kappa=1$, as can
of course
also be verified
directly from (\ref{kappa}) using 
just the Killing vector.
This is dimensionless,
since the Killing vector normalized as
above has dimensions of length.
If we instead normalize the Killing vector
as $(1/x_0)(x\partial_t+t\partial_x)$,
so that it is dimensionless and $u$ has unit
Killing energy, then the surface gravity
is equal to the 2d expansion at the horizon,
$\kappa = 1/x_0$.

For the third example, we use the concept of the
2d expansion to illuminate the sense in which
a Rindler horizon with non-zero surface gravity
is an  $M\rightarrow\infty$ limit of
Schwarzschild black hole horizons.
A black hole of mass $M$ has a surface gravity $\k=1/4M$
with respect to the Killing vector normalized to unity
at spatial infinity.
If we simply take the limit of this quantity 
as the mass is increased, holding fixed the norm
of the Killing field at infinity,
the surface gravity goes to zero. 
Equivalently, the 2d expansion 
$\theta_{2d}^{\rm hor}$ measured by
free fall observers dropped from rest at infinity
goes to zero in this limit.
To approach the Rindler limit with a non-zero 2d expansion, 
we can instead drop the 
free fall observers from a fixed proper distance
$d_0$ above the horizon, measured on a spatial
slice orthogonal to the Killing vector. 
According to (\ref{kappatheta}) this 
yields $\theta_{2d}^{\rm hor}=1/4M\sqrt{1-2M/r_0}$,
where $r_0(M,d_0)$ is the radial coordinate corresponding to the
proper distance $d_0$.
If the Killing vector is normalized at the
drop point, these observers have unit Killing
energy and therefore $\theta_{2d}^{\rm hor}$
is equal to the surface gravity.  
If $M$ is much larger than $d_0$, one finds
$r_0\approx 2M + d_0^2/8M$.
In the infinite mass limit 
$\theta_{2d}^{\rm hor}$ thus tends to $1/d_0$.
This is precisely the result, discussed in the 
previous example, for the case when the observers
are ``dropped" from a proper distance $d_0$ above a Rindler horizon. 

As a final example, we consider the 2d-expansion
in a region where the Killing vector is spacelike.
In that case the timelike geodesics can be chosen orthogonal
to the Killing vector, $u\cdot\chi=0$, so the local 
expansion (\ref{Du}) takes the form 
\beq\label{H}
\theta_{2d}  =u^a\nab_a\ln|\chi|,
\eeq
where $|\chi|=(-\chi^a\chi_a)^{1/2}$ is the 
norm of the Killing vector.
Since neighboring geodesics in the congruence 
are connected by a fixed Killing parameter,
they are separated by a proper distance $\d L$ 
proportional to $|\chi|$.
Therefore the expansion is also equal 
to $u^a\nab_a(\ln \d L)$, which is the
fractional rate of expansion in the 
Killing direction.  In a homogeneous,
isotropic cosmological metric this 2d expansion
is nothing but
the Hubble constant, provided $u^a$ is orthogonal
to all of the spatial Killing vectors. 
In the presence of anisotropy
this yields the different expansion rates along the
different Killing directions. The construction  
can even be applied
to the Killing vector $\del_t$
in the ergoregion of the Kerr metric, where it
yields a local notion of 2d Killing expansion,
given a choice of $u^a$.

 \section*{Acknowledgements}
The work of TJ was 
supported in part by the National Science Foundation under grant PHY-0601800. 
TJ is grateful to the LPT at Paris-Sud 11 for hospitality while this paper 
was being completed.

\end{document}